\documentclass[conference]{IEEEtran}

\ifCLASSINFOpdf
\else
\fi
\usepackage{amsmath}
\usepackage{mathtools}

\usepackage{hyperref}

\usepackage{xcolor}
\usepackage{graphicx}

\begin{document}

\title{Context Aware Family Dynamics based Internet of Things Access Control Towards Better Child Safety}

\author{\IEEEauthorblockN{Yasar Majib}
\IEEEauthorblockA{Cardiff University, United Kingdom\\
MajibY@cardiff.ac.uk}
\and
\IEEEauthorblockN{Charith Perera}
\IEEEauthorblockA{Cardiff University, United Kingdom\\
charith.perera@ieee.org}}

\maketitle

\begin{abstract}
Today, children are increasingly connected to the Internet and consume content and services through various means. It has been a challenge for less tech-savvy parents to protect children from harmful content and services. Internet of Things (IoT) has made the situation much worse as IoT devices allow children to connect to the Internet in novel ways (e.g., connected refrigerators, TVs, and so on). In this paper, we propose mySafeHome, an approach which utilises family dynamics to provide a more natural, and intuitive access control mechanism to protect children from harmful content and services in the context of IoT. In mySafeHome, access control dynamically adapts based on the physical distance between family members. For example, a particular type of content can only be consumed, through TV, by children if the parents are in the same room (or hearing distance). mySafeHome allows parents to assess a given content by themselves. Our approach also aims to create granular levels of access control (e.g., block / limit certain content, features, services, on certain devices when the parents are not in the vicinity). We developed a prototype using OpenHAB and several smart home devices to demonstrate the proposed approach. We believe that our approach also facilitates the creation of better relationships between family members. A demo can be viewed here: \url{http://safehome.technology/demo}.


\end{abstract}

\IEEEpeerreviewmaketitle

\section{Introduction}

Internet is growing with with enormous pace with incorporation of technologies such as Internet of Things (IoT). With this rapid growth, individuals with malicious intents are also becoming more resourceful in exploiting peoples' lives and gaining control of physical resources over cyberspace. Children are increasingly connected to the Internet and consume content and services through various means. It has been a challenge for less tech-savvy parents to protect children from harmful content and services. Children, can be misled, exploited, exposed to explicit/malicious material and can be used, by malicious parties, to gain access to their privacy and their guardians' possessions as the internet usage in children is also growing rapidly with the overall growth of internet. Extensive Internet usage is becoming an addiction for young children which is a reason for several serious mental and physical diseases like Nam and Hwang \cite{Nam2018}. Children are exposed to various risks like cyber-bullying, sexting and sexual harassment, online pornography, sexual solicitation online, Radicalization, children’s involvement in hacking/cyber-crime by using internet Livingstone et al. \cite{Livingstone2017}.


Like old days, before the internet evolved this much, when parents use to tell their children about \textit{``stander danger''}, the idea was to warn children about strangers can be potentially dangerous.  After the evolution of internet it seems that the ``stranger danger'' idea is lost, but the potential danger of strangers is still there, hidden on the other side of internet which can potentially cause serious damage to child health and privacy. Because children are sensitive by nature and can be exploited easily (with a long term effect on personality), the United Nations Convention on the Rights of the Child (UNCRC) \cite{DSa1993} approved laws for protection of children taking due account of the importance of the traditions and cultural values of each people for the protection and harmonious development of the child. UNCRC \cite{DSa1993} article 34 states that "\textit{Governments must protect children from all forms of sexual abuse and exploitation.}", article 36 states that "\textit{Governments must protect children from all other forms of exploitation, for example the exploitation of children for political activities, by the media or for medical research.}", article 41 states that "\textit{If a country has laws and standards that go further than the present Convention, then the country must keep these laws.}".  

Our proposed solution is focused on ensuring that the children are safe from the negative side of internet and getting benefited from the positive side of internet. It will play a vital role in ensuing mental and physical health of children which is at stake right now. It will also give confidence and peace of mind to guardians about the activities of their children and it will enables better relationships between family members. A research conducted by Lupianez-Villanueva \cite{Lupianez-Villanueva2016} which shows that strict regulations for businesses and better parental control software would contribute a lot towards safer and more effective use of internet by children.

In this paper, we propose mySafeHome, which utilise family dynamics to provide a more natural, and intuitive access control mechanism to protect children from harmful content and services. mySafeHome detects the presence and physical distance of two devices using Received Signal Strength Indicator (RSSI) level which is received by the wireless access point from its associated stations on every network activity. It is not only the RSSI level which can serve the purpose because of its fluctuation and variation depending on the physical environment and design of the building. There should be a system which can work dynamically using RSSI levels ignoring its fluctuation. To overcome the fluctuation problem of RSSI levels, we developed a  method to predict or detect the presence or closeness by using Machine Learning (ML). 

\textbf{Contribution:} The primary contributions and the scope of this paper are summarised below:
\begin{itemize}
\item We propose to utilise family dynamics to provide a more natural, and intuitive access control mechanism to protect children from harmful content and services.

\item We have developed a intelligent system  that can intelligently detect the existence of an adult guardian near a child or an IoT in a home environment by using the RSSI levels and other data sources.

\item We also developed home-level firewall which can dynamically control the access to internet for all types of devices (mobile/tablet/laptop/IoTs).

\item Our proposed solution based on RSSI level which works on first layers (physical layer) of OSI model for detecting physical distance. 

\item We have categorised the distance in multiple scenarios instead of binary. Our approach is to provide a sense of certainty about the internet related activities of children based on guardian's presence in different places/scenarios.

\end{itemize}
\section{Problem Formulation}

Research show that that internet usage by children has negative effects like addiction, emotional or physiological issues. Nam and Hwang \cite{Nam2018} researched previous studies conducted regarding internet usage by children and concluded that there is an increased risk of depression, social isolation, impulse control disorder, obesity, attention deficit hyperactivity disorder in children who overuse the Internet. Greenfield \cite{Greenfield2017} concluded that pathological technology use causes negative sequelae in many functional life spheres, including family life, academic and work performance, social relationship, physical health, safety and loss of sleep. According to Strasburger and Hogan \cite{Strasburger2013} the evidence is clear that ``new media'' including cell phones, iPads, and social media can contribute substantially to many different risks and health problems. Santisteban et al. \cite{DeSantisteban2018} interviewed men convicted of online grooming and shows how aggressors actively study the structural environment, the needs and vulnerabilities of the minors. 

In Santisteban et al. \cite{DeSantisteban2018}, it was referred that ICTs enables greater accessibility to children to participate in a normalized way in the virtual environment which motivated adults sees as increased opportunity for interaction with minors as the ability to operate in multiple scenarios with different potential victims simultaneously. Strasburger and Hogan \cite{Strasburger2013} recommended that parents should limit the amount of total entertainment screen time to 1 to 2 hours per day, Strasburger and Hogan \cite{Strasburger2013} also recommended that parents should monitor what media their children are using and accessing, including any websites and social media. This leave us with the question that how can we control the access to internet dynamically so the parent is aware of children's activities on internet and internet should be controlled/scheduled as per parent's preferences. 

There are several possible scenarios where this can be implemented. Some of these use-case scenarios are discussed in Table \ref{tab:1} and a demo of these scenarios are recorded which can be observed in the video.

\begin{table*}[t]
	  \caption{Usecase Scenarios}
  \centering
  \scriptsize  
  \begin{tabular}{lp{7cm}p{8cm}}
    No & Scenario Description & Factors that can be used to identify the scenarios presented in scenario  column\\ \hline
    1 & Child watching TV while parent not at home & 
    Guardian's calendar events query \\
    & & Unavailability of GD at home (using signal strength)\\
    2 & Child watching TV in the TV lounge while parent in their room &
    Light/motion sensor in guardian's room \\
    & & GD is in guardian's room (using signal strength)\\ 
    & & Any other IoT under use in guardian's room \\
    3 & Child watching TV along with parents in TV lounge & 
    GD is in the TV lounge (using signal strength) \\
    & & Any other IoT device under use in TV lounge or kitchen area which is\\
    & & categorized as guardian's use only \\
    4 & Child using mobile/tablet device while parent not at home & 
    Guardian's calendar event query \\
    & & Unavailability of GD at home (using signal strength) \\
    5 & Child using mobile/tablet in his/her room without guardian &
    Light/motion sensor in guardian's room \\
    & & GD is in guardian's room (using signal strength) \\
    & & Any other GD being use in guardian's room \\
    & & Difference in signal levels of CAD and GD \\
    6 & Child operating mobile/tablet with guardian around (same room / area) & 
    Similarity of signal strength of both child's and parent's devices \\
  \end{tabular}

  \label{tab:1}
\end{table*}

\section{System Architecture}

\begin{figure*}[!ht]
	\centering
	\includegraphics[width=1.7\columnwidth]{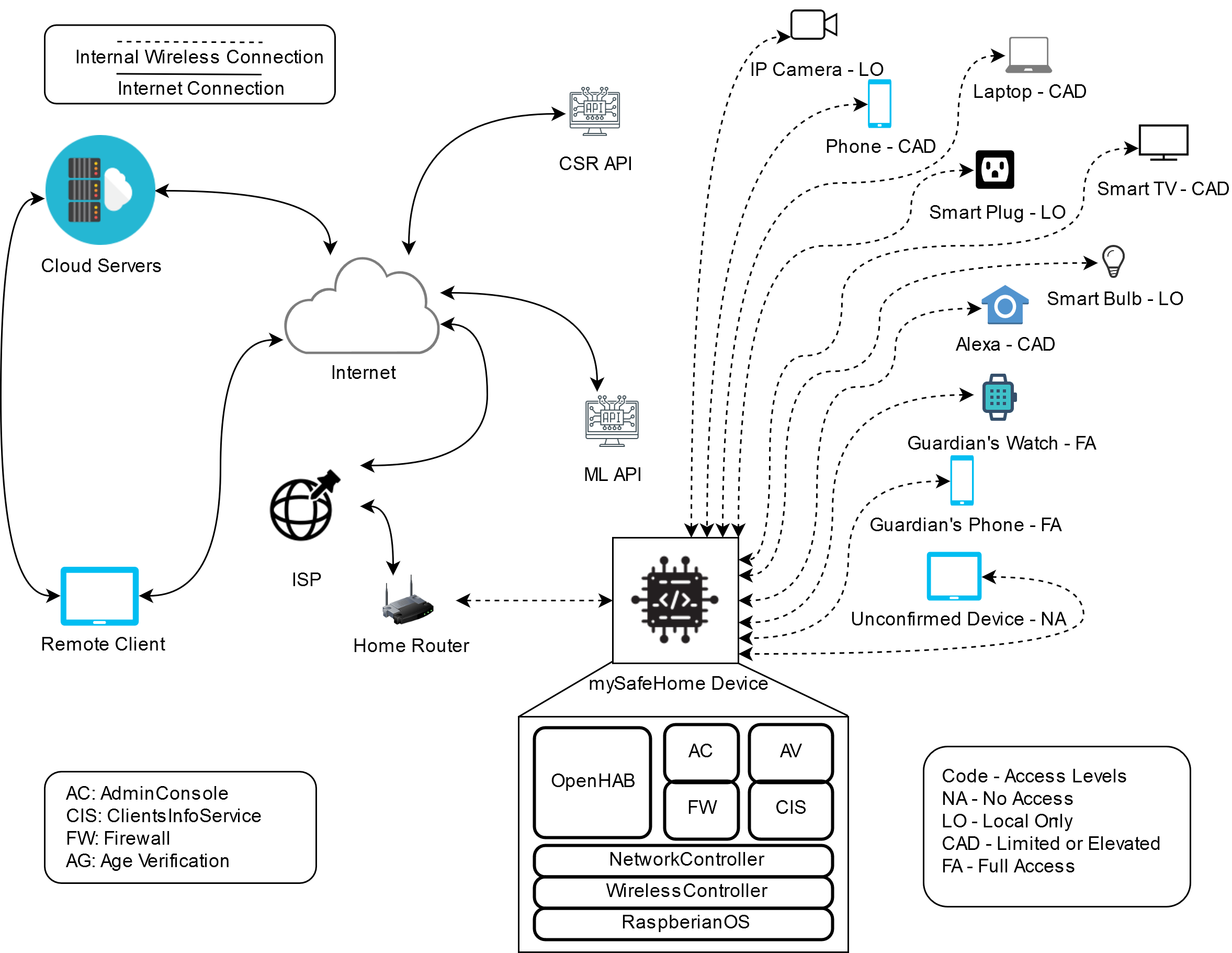}
	\caption{System Architecture of mySafeHome}
	\label{fig:SystemArchitecture}
	\vspace{-15pt}
\end{figure*}

The system depends on various entities e.g. CADs, GDs, IoTs, mySafeHome, Home Router and Cloud Services as presented in Figure \ref{fig:SystemArchitecture}. In mySafeHome, access control dynamically adapts based on the physical distance between family members. We used a Raspberry Pi 3B+ (RPi) device with Raspbian OS and on top of that we install and configured existing networking services such as \texttt{DHCP Server}, \texttt{DNS Server}, \texttt{IPTables}, \texttt{IPSets} along-with our custom developed services like \texttt{AdminConsole}, \texttt{ClientsInfoService}, \texttt{MLService} as shown in Figure \ref{fig:SystemArchitecture}. The system is designed in a way that it can detect child-guardian and IoT-guardian physical distance. A web-based user interface is also developed for setting up, administering and monitoring purpose.  To manage IoT devices we are using OpenHAB, a cross-platform software to integrate all kinds of Smart Home technologies and devices, which is installed in the same RPi.  We have defined five different access levels to control network access of the IoT devices. We have defined key concepts used in this paper as well.
\begin{itemize}
\item \textbf{No Access:} Deny access to newly added unconfirmed devices by default.
\item \textbf{Local Access Only}: Actuators or sensors which do not require internet access.
\item \textbf{Limited Internet Access}: Access to limited websites for CADs in case of unavailability or parent or as configured.
\item \textbf{Elevated Internet Access}: DNS based filtered access to internet for CADs in case of availability or guardian or as configured.
\item \textbf{Full Access}: Unrestricted internet access for guardians. 
\end{itemize}

\textbf{Actors}
\begin{itemize}
\item \textit{Guardian:} A person who administrator of the home, the system will consider device as administrator when its access level is Full Access.
\item \textit{Child:} All the minors at home who has, as preferred by guardian, filtered internet access. Devices with internet access which can be accessible by child will have either Limited Internet Access or Elevated Internet Access. 
\end{itemize}

\textbf{System Components}
\begin{itemize}
\item Actuators can be Smart Lights and Smart Plugs in our scenarios, access level for these devices are Local Only as most of the actuators in our system will not require internet access to operate.
\item Child Accessible Devices (CADs): refers to a device which can be accessible by a child e.g. Smart TV, Mobile, etc. Each device will be scanned and evaluated for access level in the background.
\item Guardian's Devices (GDs): refers to a device under use by a guardian. The system will look for all GDs for each CAD to check detect physical distance.
\end{itemize}
We used following factors for  dynamic access control:
\begin{itemize}
\item Guardian's Calendar Event: An external calendar which serves the events list of guardian, in our system we are using \texttt{Google Calendar} via \texttt{google-python-api}.
\item Physical Distance using RSSI: We are using RSSI levels to determine the physical distance of two devices.
\item Schedule: The system allowed guardian to set schedule to allow internet access in preferred time ranges. 
\end{itemize}
\section{Implementation}
\begin{figure}[h]
	\centering
	\includegraphics[scale=0.55]{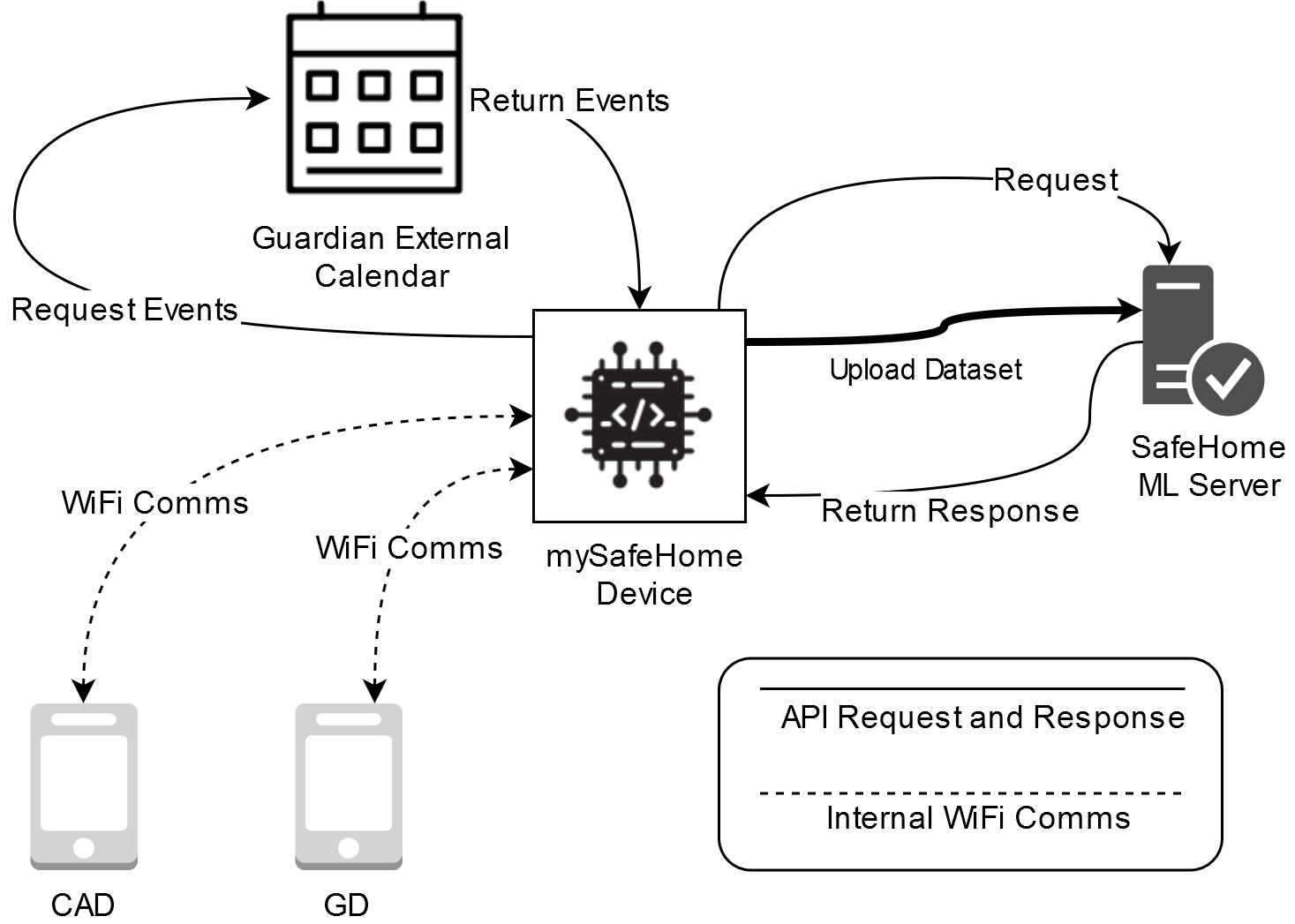}
	\caption{Dynamic Access Control}
	\label{fig:MLProcess}
\end{figure}

To make our "mySafeHome Device" as shown in Figure \ref{fig:MLProcess} work as a router we, at first, installed and configured all necessary services like \texttt{DHCP}, \texttt{DNS}, \texttt{HostAPD}, \texttt{IPTables} and OpenHAB. We configured two wireless network interfaces for handling internal and external traffic separately. We used RaspbianOS to deploy all services. Once all services are working normal the devices at home will be connected to internal wireless access point of RPI, it will become gateway for all devices. The RPI can get RSSI level of every connected device and can also control the network access of each device.\\
\textbf{Dataset for Machine Learning}: Next step is to generate dataset for machine leaning, as our dataset is based on RSSI levels, we connected two mobile devices (Dev1, Dev2) to RPI's wireless network and started the process of creating dataset as shown in Figure \ref{fig:MLProcess}. To read RSSI levels from wireless handler we used \texttt{iw dev phy INTERFACE-NAME station dump} command, and parse output to python script which converts it into array. Following equation can represent the operation:\texttt{Y = [(RSSI(i)), Label]}. The process repeats \texttt{X} times, \texttt{X} is sensitivity which used to ignore the fluctuation effect of RSSI levels. In above mentioned formula \texttt{i=[0, 1, 2, 3, .... X*2]} where \texttt{even(i)} represents Dev1 and \texttt{odd(i)} means Dev2. Label represents Boolean value of closeness of devices True means near and far/away means False. As we need to keep privacy issue under consideration, we remove identification of devices from the dataset so it contains only RSSI levels of devices.\\
\textbf{Dynamic Access Control}: There are several processes need to be running in the background to enable dynamic access control, these processes are designed to interact with user as well as to check for all factors and make decisions. We have developed couple of services which can run on RPI to serve the purpose. These services interact with user and/or several sources like database, OS files and web-based APIs.
\\
\texttt{ClientsInfoService}, the core service running in background, does not required any user interface as it is solely depending on information stored in sources. First step of this service is to initialize SQL database, read DHCP server's lease list and add client devices into database table. Secondly, it will start evaluating  all factors of access control for each CAD to control its access to internet. 
\begin{itemize}
\item \textit{Schedule:} is optional which has to be configured by guardian to enable access to internet on the preferred times of the day depending on other factors. If schedule is not configured, it will allow access 24/7.
\item \textit{Google Calendar API:} We integrated calendar API to detect if there is any event exists for current time, if so then the guardian is considered to be out of home. In that case our system will limit internet access of all CADs.
Machine Learning Algorithm: The most critical part of this service is to check if the GD is physically close to CAD. Following are the steps of the process:
\item \textit{Get RSSI levels of CADs and GDs:} The system will query all CADs and GDs from the database, it will make sets of each CAD with all GDs (one-by-one). Then read RSSI levels from wireless handler for \texttt{X} number of times, \texttt{X} defines sensitivity. We use \texttt{X} to ignore the fluctuation in value of RSSI levels. Once the RSSI data of devices is stored in local arrays, it will remove identification of devices from the data i.e. MAC Address.\\
\texttt{{Z = [(CAD RSSI, GD RSSI) * X]}}\\
The set \texttt{Z} will be sent to machine learning web-api where it will be processed, and result will be returned as Boolean. The system will limit/elevate internet access of CAD according to the result. The service will update database on every update so the guardian can view real-time status of devices using web-based interface.\\
The other service, \texttt{AdminConsole}, allows guardian/owner to interact with the system in order to complete the initial setup and later administer/monitor the devices. Guardian can view/connect available wireless networks, setup internal wireless access point, view/change access control of devices, restore device to default settings, restart or shutdown device, etc. by using this interface. This interface accessible via device's touch screen (kiosk mode) and/or using web-browser in any other device.
\end{itemize}
\section{Evaluation and Demonstration}
We evaluated the device's functionalities in Table \ref{tab:1} scenarios as shown in Figure 3. Circle means that child is watching TV and diamond means child is using a CAD. Red color shows restricted/limited access, blue represents device specific restrictions and green means elevated internet access or full operations. Evaluation of the device's functionalities for the given scenarios in mentioned in Table 1 need pre-requisite steps to be completed. \textbf{Power-on Device:} Unpack and power-on device assumed that device's current settings are default. \textbf{Device Initialization:} Connect to internet using available wireless network, setup internal access point to allow devices to connect to RPI, setup guardian's device, setup CADs.
\begin{figure*}[!ht]
	\centering
	\includegraphics[scale=0.15]{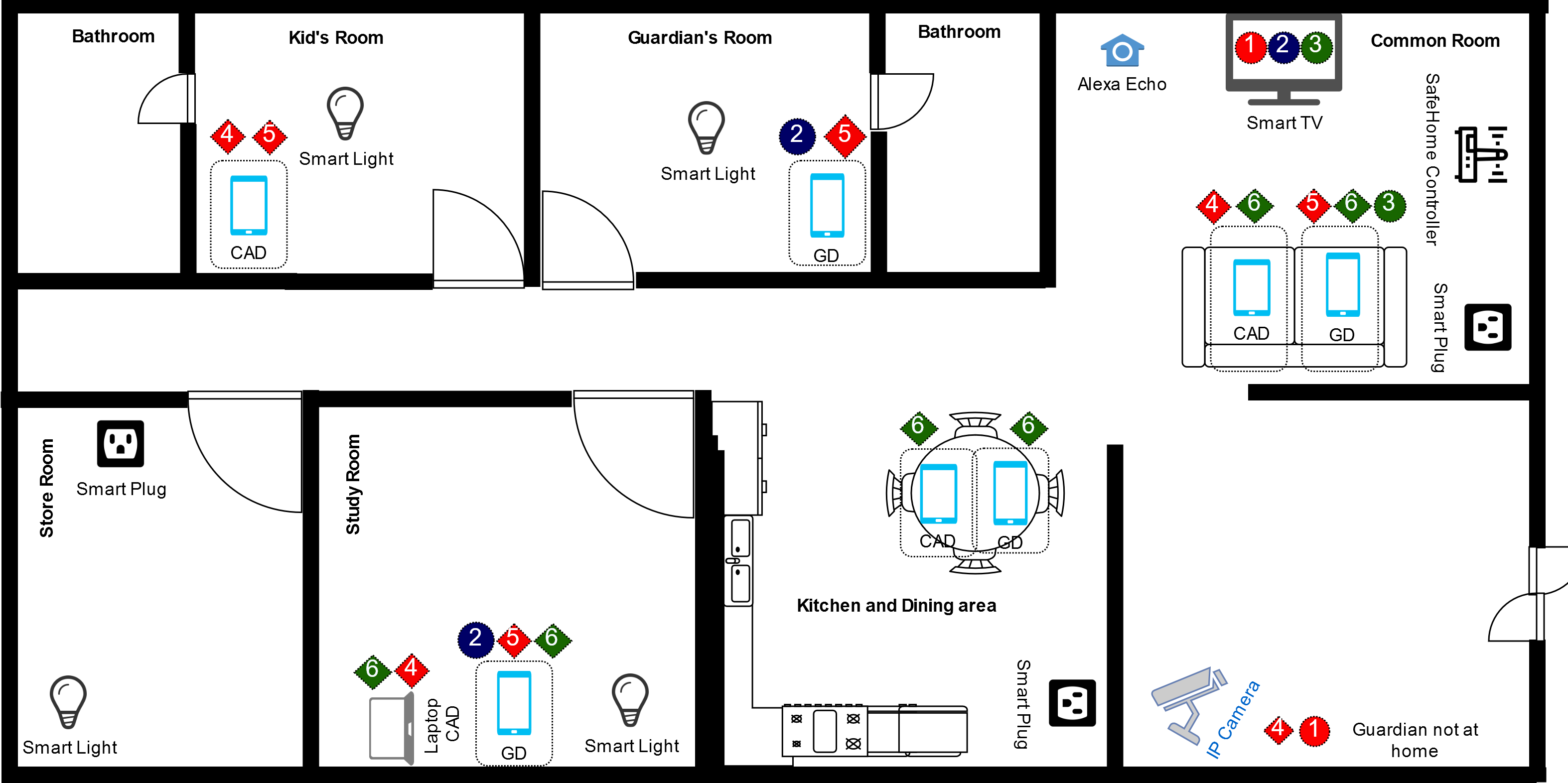}
	\caption{Demo Scenarios}
	\label{fig:Scenarios}
	\vspace{-10pt}
\end{figure*}

\begin{itemize}
\item\textbf{Scenario 1}: Successfully detected using External Calendar API that guardian is not at home so access of Smart TV to internet is set limited to certain ``safe for kids'' websites only.
\item\textbf{Scenario 2}: Successfully detected using RSSI level that GD is far from Smart TV (by examining difference between RSSI levels of GD and Smart TV) but present in home. Smart TV volume is increased and locked.
\item\textbf{Scenario 3}: Successfully detected using External Calendar API and RSSI level that GD is near to Smart TV and unlocked its volume.
\item\textbf{Scenario 4}: Successfully detected using External Calendar API that guardian is not available at home to limit internet access of CADs.
\item\textbf{Scenario 5}: Successfully detected using RSSI level that GD is far from CAD but available at home to limit its access to internet.
\item\textbf{Scenario 6}: Successfully detected using RSSI level that GD is near to CAD to elevate its access to internet.

\end{itemize}
\section{Related Work}
There has been significant research being done related to IoT based smart homes, its benefits, its challenges and possible options in the recent years. We briefly review three related areas, namely, home automation, home activity detection, and RSSI based Location awareness.


\textbf{Smart Home automation using central controller devices:} There is a comprehensive amount of work done on controlling home equipment in a smart home using central hubs like the system proposed by Saputri and Rofiq \cite{Saputri2018} who developed a web-based system using Raspberry Pi with python code to control the home equipment over the internet but the solution is focused on reducing waste of electricity by turning on/off the equipment. Khan et al. \cite{Khan2016} proposed a smart home control system to mitigate the effect of interference and reduces the energy consumption of the smart home appliances. Khan et al. \cite{Khan2016} proposed smart light control system is tunes the illumination level in a room by incorporating the natural light. Khan et al. \cite{Khan2016} also proposed a management station with the system that is designed to control the working time of the smart home appliances, from Khan's work we can drive that a system can be built that can instruct IoTs to perform certain task depending on the state of other sensors or IoTs. A similar work done by Singh et al. \cite{Singh2018} in which a solution which controls some home appliances like light, fan, etc. using various sensors like LM35, IR, LDR, Node MCU ESP8266 and Arduino UNO. Singh et al. \cite{Singh2018} proposed solution also detects the presence or absence of human object in the house using these sensors, the reason Singh et al. \cite{Singh2018} is not using radio signals for the purpose is, which in fact is true, that they can be easily intercepted and prone to distortions but since almost every media oriented IoT device, in our case, now days is connected to internet using Wi-Fi that makes it necessary for us to rely on radio signals as the primary detection and communication mechanism. Zhong and Hu \cite{Zhong2013} purposes use of Intelligent Decision Support System to improve the intelligence of smart home system but the purposed system is presented to control the output functions of IoT devices only and the system required user request for every event.

\textbf{Human Activity detection in a Smart Home:} Several researches have been conducted to detect activity based on the information received by IoT and other devices like mobile phones. Kashimoto et al. \cite{Kashimoto2016} propose a activity recognition method using energy harvesting PIR and door sensors, this approach can be used to categorize general activities and presence on human object in a certain area of the house but may not be able to identify the object considering the fact that we want to calculate distance between two human objects. Another proposed a data-driven solution by Kodeswaran et al. \cite{Kodeswaran2016a} to detect activities of daily living in IoT enabled smarthomes, these activities are also categories as general activities and can not confirm the identity of human object in a smart home. Dang et al. \cite{Dang2018} proposed an idea for detecting human activities, based on the odors generated by these activities, using an electronic nose based on many sensors connected with raspberry pi. Dang et al. \cite{Dang2018}'s idea is limited to detecting certain type of activities but indoor location or distance of objects can not be determined.

\textbf{RSSI based Location awareness} There is also some work being carried out to detect anomalies using RSSI associated to the wireless transmissions of connected devices in a home, like Roux et al. \cite{Roux2017} presented solutions based on profiling of RSSI, Roux et al. \cite{Roux2017} used a machine learning algorithm to characterize legitimate communication and to identify suspicious scenarios. Roux et al. \cite{Roux2017} work is focused on intrusion detection, in addition to Roux et al. \cite{Roux2017}, our work is using machine learning algorithm to detect closeness of two Wi-Fi devices which are communicating with a single wireless access point in an indoor place. According to Zanca et al. \cite{Zanca2008} the indoor radio channel is very predictable so he suggested use of time-to-flight of pressure waves, or ancillary radio hardware like multiple and/or directional antennas. Zanca et al. \cite{Zanca2008} also added that the alternative solutions will require more energy, dedicated hardware resulting making it expensive. We agree with Zanca et al. \cite{Zanca2008} about the unpredictability that is the reason of selecting machine learning for RSSI data received in a specific time frame to predict the result. Pathak et al. \cite{Pathak2014} also did similar work by setting up three wireless access points and a mobile application to scan RSSI levels and send central server, the accuracy rate might be better but it is not practical in case of heterogeneous devices' network and a we can not increase cost of adding more WAPs in smart homes. Furst et al. \cite{Furst2018} proposed the usage of visual 3D models for localization and distance measurement with high accuracy, where he mentioned that localization systems often suffer from large data collection and calibration overhead in new environment and RSSI fingerprinting based localization systems require fingerprinting database, as our focus in only to determine physical distance of two devices, we only need RSSI data to feed to machine learning for better accuracy.

Despite of the identified problems of excessive internet usage and availability of extensive research in the area smarthomes, distance measurement using physical layer data there is no such solution, as per best of our knowledge, that can seamlessly control internet access of child device or an IoT dynamically depending on the presence or absence of a guardian from the scene.

\section{Conclusion and Future Work}
We utilised family dynamics to provide a more natural, and intuitive access control mechanism to protect children from experiencing harmful content and services over internet. Our access control system dynamically adapts based on the physical distance between family members. We used combination of legacy systems, physical layer data as well as third-party APIs for detection. Our developed system does not depend on the software system of IoTs rather it can work with almost any IoT which uses WiFi connection. We have successfully developed a prototype using OpenHAB and several smart home devices to demonstrate our approach. Our approach can facilitates the creation of better relationships between family members. 
 
Our proposed system is currently using fixed sensitivity level and works with two devices at a time but it can be expanded to multiple devices simultaneously and with variable sensitivity according to the physical environment. Evaluation of the purposed system need to be done in natural/real scenarios so that it will evolve as a mature system to be adapted by people at large in the future use. It can also be incorporated/embedded in home routers, provided by ISPs, so the features can be utilized by default without adding additional devices. The accuracy of output is dependent on the diversity of dataset for machine learning, better algorithms can also be formed using this novel approach.

\section*{Acknowledgement}

We acknowledge the support by PETRAS2 EP/S035362/1 and EPSRC Capital Award  for Early Career Researchers.

%



%

\bibliographystyle{abbrv}
\bibliography{library}

\end{document}